\documentclass[12pt]{article}
\usepackage[utf8]{inputenc}
\usepackage{graphicx}
\usepackage{bm}
\usepackage{amsmath}
\usepackage{float}
\usepackage{caption}
\usepackage{subcaption}
\usepackage{multirow}
\usepackage{amsfonts}
\usepackage{textcomp, gensymb}
\usepackage{algpseudocode}
\usepackage{algorithm}
\usepackage{textcomp}
\usepackage[top=0.8in,bottom=0.8in,left=0.8in,right=0.8in]{geometry}
\usepackage{makecell}
\usepackage{xcolor}
\usepackage{cite}
\usepackage{appendix}
\usepackage{url}
\usepackage{amsthm}
\usepackage{makecell} % Allows line breaks in cells
\usepackage{tabularx} % Add this to the preamble

\DeclareSymbolFont{matha}{OML}{txmi}{m}{it}% txfonts
\DeclareMathSymbol{\varv}{\mathord}{matha}{118}

\newcommand{\eqn}[1]{eqn.~(\ref{#1})}

\newcommand{\sect}[1]{Section~\ref{#1}}

\newcommand{\fig}[1]{Fig.~\ref{#1}}

\newcommand{\tab}[1]{Table~\ref{#1}}

\theoremstyle{definition}

\newcommand{\angstrom}{\mbox{\normalfont\AA}}

\title{Interatomic potential development for topological insulator Bi$_{1-x}$Sb$_{x}$ and its dislocation by force-following active learning}
\author{Moon-ki Choi, Daniel Palmer, and Harley T. Johnson}
\date{}
\begin{document}
\maketitle
\begin{abstract}
We introduce an force-following active learning algorithm that integrates density functional theory (DFT) with the Gaussian Approximation Potential (GAP) framework to develop a robust interatomic potential (IP) for a dislocation in a topological insulator, Bi$_{1-x}$Sb$_x$. Starting from an initial potential, IP$_0$, trained on unit cell data from strained Bi-Sb binaries, our active learning approach iteratively refines the IP during a structural relaxation. In each cycle, if the force error (uncertainty) of any atom near the dislocation core exceeds a threshold value, the IP$_i$ is efficiently retrained (IP$_i\rightarrow$ IP$_{i+1}$) by incorporating DFT-computed forces and energies of atoms near the high-uncertainty atom. This strategy ensures that the relaxation process maintains a low force error until full convergence is achieved. Consequently, the final IP, here IP$_5$, has two capabilities: (1) it reproduces the relaxation pathway observed during the active learning process unlike the initial IP$_0$, which lacks prior dislocation core knowledge; and (2) it capture the lattice and elastic properties of Bi-Sb binaries across a range of Sb concentrations. We also evaluate dislocation properties (Peierls stresses and dislocation generation by compression) to assess the performance of the trained potential, IP$_5$.
\end{abstract}

\section{Introduction}
The rapid advancement of data-driven interatomic potentials (IPs) has gained attention with the integration of machine learning, leveraging the Universal Approximation Theorem through neural networks \cite{cybenko1989,nielsen2015,hornik1989}. This integration has led to the development of highly accurate and generalizable IPs, such as the universal IP \cite{chen2022}, equivariant graph neural network IP, NequIP \cite{batzner20223}, and Gaussian Approximation Potential (GAP), which  provides atomic forces and energies while quantifying uncertainties \cite{bartok2010}. These developments have been also accelerated by massive quantum simulation data archived by resources such as the Materials Project \cite{jain2013}, AFLOW \cite{curtarolo2012}, and ColabFit \cite{vita2023}. With the advent of accessible tools for developing IPs using big data, new classes of materials and their crystalline defects, which would have been impossible to model only a few years ago, become promising new targets for atomistic simulation.  

As an example, we consider the particularly intriguing application area of the mechanics of topological insulators (TIs). In these materials, a dislocation with a specific Burgers vector may induce a gapless (i.e., metallic) state that facilitates electron transport through the dislocation \cite{hasan2010,ran2009}. 
Theoretically, this gapless state is formed due to a dislocation if the Burgers vector of the dislocation ($\mathbf{b}$) satisfies the following: 
\begin{gather}\label{eqn:dis_TI}
\mathbf{b}\cdot\mathbf{M}=\pi\;(\mathrm{mod}\;2\pi), \\
\mathbf{M}=\frac{1}{2}(v_1\mathbf{G}_1+v_2\mathbf{G}_2+v_3\mathbf{G}_3),
\end{gather}
where $(v_1,v_2,v_3)$ are three $Z_2$ invariants $(v_0;v_1,v_2,v_3)$ of TI \cite{fu2007}, and the vectors $\mathbf{G}_1$, $\mathbf{G}_2$, $\mathbf{G}_3$ are the primitive reciprocal lattice vectors. One example of these TIs is the Bi-Sb binary, Bi$_{1-x}$Sb$_x$. Bi$_{1-x}$Sb$_x$ becomes a strong topological insulator in the 
range $0.07<x<0.22$ with invariant $(1;111)$.  For Bi$_{1-x}$Sb$_x$ in the range of $0.07<x<0.22$, $\mathbf{b}=\langle 100\rangle$ satisfies \eqn{eqn:dis_TI} \cite{tokumoto2019}. Therefore, understanding the atomistic mechanics of the dislocation in this material is crucial for elucidating the interplay between structural defects (i.e., dislocation) and the topological properties of the material. Density Functional Theory (DFT) is a viable option for studying dislocation mechanics. However, running DFT simulations for a dislocation requires substantial computational resources due to a large atomistics unit cells needed to accurately represent the dislocation.  While interatomic potentials can manage these large systems, none has yet been specifically developed for the Bi-Sb binary systems, nor for their dislocations. To address similar challenges in other materials systems, previous research has explored active learning algorithms for applications such as crack propagation, screw dislocation, and crystal structure prediction \cite{hodapp2020,zhang2023,podryabinkin2019,jinnouchi2020}. These studies benefit from the uncertainty quantification capabilities of IP such as GAP or the moment tensor potential. These active learning strategies enable more efficient, targeted simulations, thereby facilitating the development of efficient interatomic potentials for complex systems.

In this paper, we develop and employ IP for Bi-Sb binary systems (Bi, Bi$_7$Sb$_1$, Bi$_3$Sb$_1$, BiSb, and Sb structures), with a particular focus on dislocation properties in the Bi$_7$Sb$_1$ ($0.07<x<0.22$ with $x=0.125$) system. First, we generate force and energy data of strained and perturbed  Bi-Sb binaries using DFT calculation. The relationship between the rhombohedral and monoclinic structure of Bi$_7$Sb$_1$ is discussed in \sect{sec:method:atom_Bi7Sb1}. Next, we introduce a dislocation with Burgers vector $\mathbf{b}=[100]$ into the Bi$_7$Sb$_1$ atom structure under fixed boundary conditions (\sect{sec:method:dis_model}). An active learning algorithm, detailed in \sect{sec:method:active}, is developed to iteratively train a GAP that captures the relaxation of the dislocation, (\sect{sec:result:relaxation}) guided by force error. After training the GAP for the dislocation system using the active learning process, we assess the performance of the trained IP by simulating Peierls stress (\sect{sec:result:peierls}) and and dislocation generation by compression of Bi$_7$Sb$_1$ (\sect{sec:result:compression}).

\section{Method}\label{sec:method}
\subsection{DFT calculation for data set}\label{sec:method:DFT}
All DFT calculations in this work were performed using the Quantum ESPRESSO (QE) software package \cite{QE-2017}. In QE, the Generalized Gradient Approximation (GGA) with the Perdew-Burke-Ernzerhof (PBE) exchange-correlation functional was used. Norm-conserving pseudopotentials, based on GGA+PBE, were used for all calculations. The plane-wave kinetic energy cutoff for the wave functions was set to 50 Ry, while a cutoff of 500 Ry was employed for the charge density. Brillouin zone integrations were carried out using a $k$-point spacing of 0.02 $\angstrom^{-1}$. Self-consistent field iterations continued until the total energy converged to within 1 meV per atom. Additionally, spin-orbit coupling (SOC) was enabled for all calculations, as it is crucial for accurately capturing the elastic properties of Bi-Sb binary systems \cite{singh2018}.

Rhombohedral structures of Bi, Bi$_3$Sb$_1$, BiSb, and Sb were obtained from the Materials Project \cite{jain2013}, while the structure of monoclinic Bi$_7$Sb$_1$ was taken from \cite{singh2016}. These five structures were fully relaxed with the DFT calculation framework described above, with both the atomic positions and the simulation cell allowed to relax until the forces and final stress in each system reached zero.
After the relaxation, we systematically applied six independent strain components $\epsilon_{xx}, \epsilon_{yy}, \epsilon_{zz}, \epsilon_{yz}, \epsilon_{xy},$ and $\epsilon_{xz}$ to the relaxed unit cells of Bi, Bi$_7$Sb$_1$, Bi$_3$Sb$_1$, BiSb, and Sb. Each strain component was randomly chosen from the discrete set $\{-0.06,-0.03,0.00,0.03,0.06\}$, with the constraint that no two strain tensors share four or more identical components.\footnote{Although one might be concerned that using discrete 
strain increments overlook intermediate strains, we confirmed the the trained GAP model with this scheme produces elastic stiffness consistent with continuously varying strains up to 6\%.}
The deformed lattice vector ($\mathbf{R}'$) is obtained via
\begin{equation}
\mathbf{R}' = \mathbf{F}\mathbf{R} = \bigl(\mathbf{I}+\mathbf{E}\bigr)\mathbf{R},
\end{equation}
where $\mathbf{F}$ is the deformation gradient, and $\mathbf{E}$ is the small strain tensor defined as:
\begin{equation}
\mathbf{E} = 
\begin{bmatrix}
\epsilon_{xx} & \tfrac{1}{2}\epsilon_{xy} & \tfrac{1}{2}\epsilon_{xz} \\
\tfrac{1}{2}\epsilon_{xy} & \epsilon_{yy} & \tfrac{1}{2}\epsilon_{yz} \\
\tfrac{1}{2}\epsilon_{xz} & \tfrac{1}{2}\epsilon_{yz} & \epsilon_{zz}
\end{bmatrix}.
\end{equation}
Following the application of strain, the $x$, $y$, and $z$ positions of each atom were perturbed randomly within a range of $\pm0.5\,\angstrom$ to capture stochastic variations in atomic positions in the relaxation.  
\begin{table}[h]
    \centering
    \footnotesize
    \begin{tabular}{cc}
    \hline\hline
        Structure & Number of DFT data \\
        \hline
        Bi &  100\\
        Bi$_{7}$Sb$_{1}$ &  200\\
        Bi$_{3}$Sb$_{1}$ &  160\\
        BiSb &  200\\
        Sb &  100\\
        Total & 760\\
    \hline\hline
    \end{tabular}
    \caption{Number of DFT data for each Bi–Sb system used in the initial GAP training.}
    \label{tab:structure_data}
\end{table}
\tab{tab:structure_data} summarizes the number of strained and perturbed DFT data points for each Bi–Sb system used in training. For each system, 10 data points were reserved for validating predicted force and energy of the GAP (see Section I in Supplementary Materials for comparison). We use one two-body descriptor and two SOAP descriptors with the different setting. Details of the GAP are described in {\color{red} \{Github page will be added\}}.
The GAP trained on this dataset is denoted as IP$_0$. IP$_0$ is used as the initial GAP model for the active learning procedure for the relaxation (\sect{sec:result:relaxation}). Note that the DFT training dataset described in \sect{sec:method:DFT} only includes strained and perturbed configurations of Bi-Sb binaries unit cells, so IP$_0$ therefore does not include information about dislocations. 

\subsection{Atomistic modeling of rhombohedral Bi$_7$Sb$_1$ with a dislocation}\label{sec:method:atom_Bi7Sb1}
Singh \cite{singh2016} predicted the atomistic structure of Bi$_7$Sb$_1$ as a monoclinic phase using the minima hopping method combined with DFT calculations. In our work, we adopt this monoclinic structure to model Bi$_7$Sb$_1$ containing an edge dislocation ($\mathbf{b}=[100]$), which is of interest because it induces a metallic state. Because  previous literature \cite{jain1959,fu2007,teo2008} generally characterizes  Bi$_{1-x}$Sb$_{x}$ as rhombohedral (or equivalently, hexagonal) phase, it is essential to investigate the monoclinic structure of Bi$_7$Sb$_1$ within this context.
\begin{figure}[ht]
    \centering
    \includegraphics[width=0.95\textwidth]{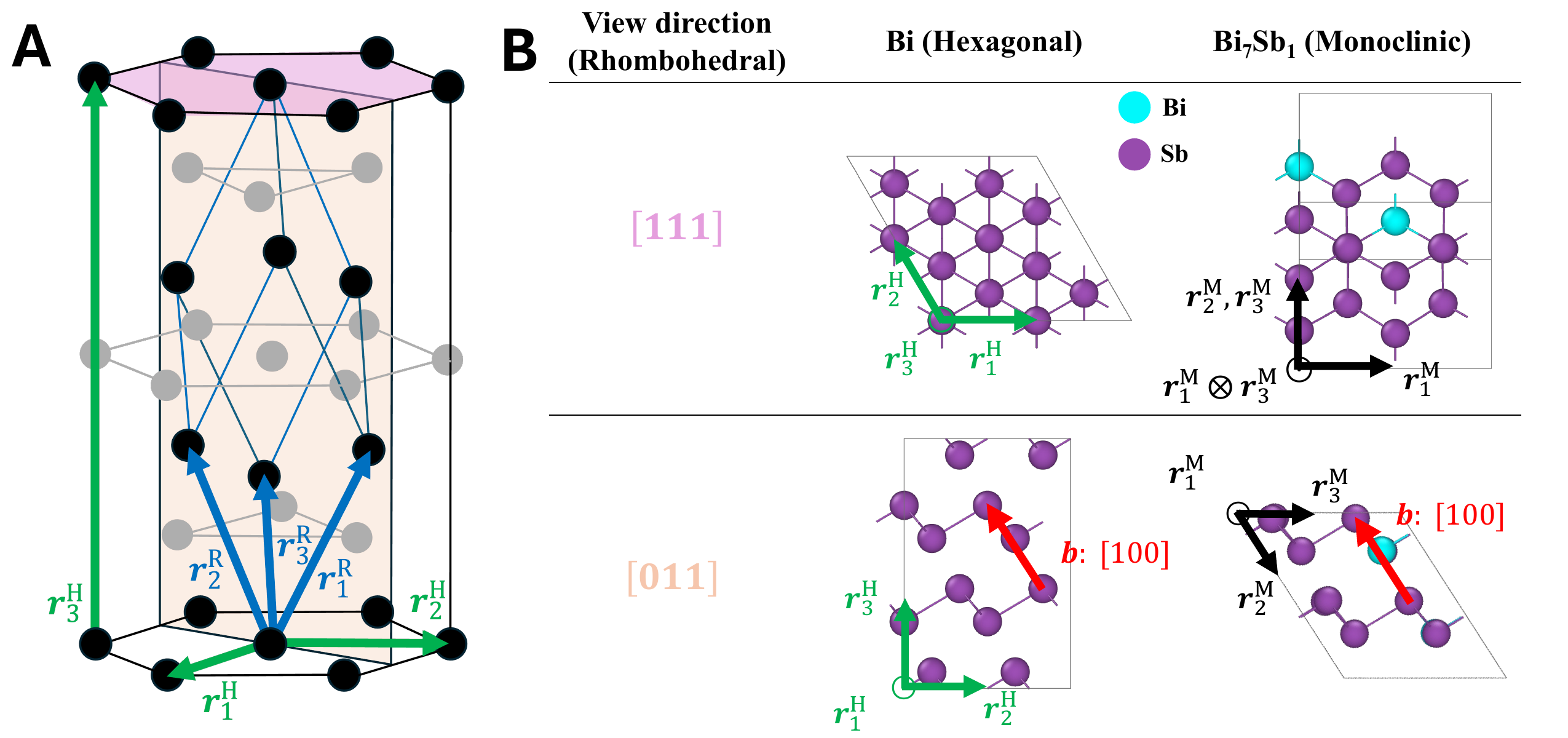}
    \caption{Structural similarity between hexagonal Bi and monoclinic Bi$_7$Sb$_1$. (A) Hexagonal and rhombohedral structure of Bi. Black and gray atoms present alternating layers. (B) The two atomistic structures above represent the viewing direction of $\mathbf{r}^{\rm H}_3$ for Hexagonal Bi and the equivalent direction ($\mathbf{r}_{2}^{\rm H} \otimes \mathbf{r}_{3}^{\rm H}$) on monoclinic Bi$_7$Sb$_1$. The two structures below represent the viewing direction of $\mathbf{r}^{\rm H}_1$ on hexagonal Bi and the equivalent direction ($\mathbf{r}_{1}^{\rm H}$) on monoclinic Bi$_7$Sb$_1$. The Burgers vector $\mathbf{b}=[100]$ in rhombohedral notation is presented as a red arrow.}
    \label{fig:atom_view}
\end{figure}
\fig{fig:atom_view} illustrates the structural similarity between hexagonal Bi and monoclinic Bi$_7$Sb$_1$. The lattice vectors of the hexagonal and rhomboehdral cells of Bi (see \fig{fig:atom_view}A)  are denoted as
\begin{gather}
\mathbf{r}{1}^{\rm H} = 4.61\,\hat{\mathbf{x}},\quad
\mathbf{r}{2}^{\rm H} = -2.3\,\hat{\mathbf{x}} + 3.99\,\hat{\mathbf{y}},\quad 
\mathbf{r}{3}^{\rm H} = 11.98\,\hat{\mathbf{z}}, \\ 
\mathbf{r}{1}^{\rm R} = -2.3\,\hat{\mathbf{x}} + 1.33\,\hat{\mathbf{y}} + 3.99\,\hat{\mathbf{z}},\quad 
\mathbf{r}{1}^{\rm R} = 2.3\,\hat{\mathbf{x}} + 1.33\,\hat{\mathbf{y}} + 3.99\,\hat{\mathbf{z}},\quad
\mathbf{r}{1}^{\rm R} = -2.66\,\hat{\mathbf{y}} + 3.99\,\hat{\mathbf{z}},
\end{gather}
while those of the monoclinic Bi$_7$Sb$_1$ cell are 
\begin{equation}\label{eqn:lat_vec_mono}
\mathbf{r}{1}^{\rm M} = 9.22\,\hat{\mathbf{x}},\quad
\mathbf{r}{2}^{\rm M} = 9.66\,\hat{\mathbf{y}},\quad
\mathbf{r}{3}^{\rm M} = 4.32\,\hat{\mathbf{y}} + 6.71\,\hat{\mathbf{z}}.
\end{equation}
\fig{fig:atom_view}B shows the atomic arrangement along the $[111]$ and $[011]$ directions (rhombohedral) for both hexagonal Bi and monoclinic Bi$_7$Sb$_1$. We found that the arrangement of atoms observed along $\mathbf{r}_{3}^{\rm H}$ in hexagonal Bi aligns with the direction defined by the cross product of $\mathbf{r}^{\rm M}_1$ and $\mathbf{r}^{\rm M}_3$ (i.e., $\mathbf{r}^{\rm M}_1 \otimes \mathbf{r}^{\rm M}_3$) in monoclinic Bi$_7$Sb$_1$. Meanwhile, for the $[011]$ direction, we found that the view along the $\mathbf{r}_1^{\rm H}$ direction in hexagonal Bi coincides with the $\mathbf{r}_1^{\rm M}$ direction in monoclinic Bi$_7$Sb$_1$.\footnote{Although the atomic arrangements in hexagonal Bi and monoclinic Bi$_7$Sb$_1$ are nearly identical, a slight difference in bond lengths is observed due to the incorporation of Sb atoms in Bi$_7$Sb$_1$. Specifically, the Bi–Sb bond length differs slightly from the Bi–Bi bond length, resulting in minor variations.} Despite Bi$_7$Sb$_1$ crystallizing in a monoclinic structure, its atomic arrangement is nearly identical to that of Bi-Sb binaries in the rhombohedral (hexagonal) cell.\footnote{We further confirmed that the structural similarity is maintained not only along the two illustrated viewing directions but also across several additional orientations.}

Using the structural similarity between hexagonal Bi and monoclinic Bi$_7$Sb$_1$ shown in \fig{fig:atom_view}, we define effective hexagonal lattice vectors ($\bar{\mathbf{r}}_{1}^{\rm H}, \bar{\mathbf{r}}_{2}^{\rm H}$, and $\bar{\mathbf{r}}_{3}^{\rm H}$) in the monoclinic  Bi$_7$Sb$_1$ structure. These lattice vectors are defined as 
\begin{equation}
\bar{\mathbf{r}}_{1}^{\rm H} = a^*\frac{\mathbf{r}_{1}^{\rm M}}{\vert\mathbf{r}_{1}^{\rm M}\vert},\quad
\bar{\mathbf{r}}_{2}^{\rm H} = a^*\frac{\mathbf{R}\mathbf{r}_{1}^{\rm M}}{\vert\mathbf{r}_{1}^{\rm M}\vert},\quad
\bar{\mathbf{r}}_{3}^{\rm M} = c^*\frac{\mathbf{r}_{1}^{\rm M} \times \mathbf{r}_{3}^{\rm M}}{\vert\mathbf{r}_{1}^{\rm M} \times \mathbf{r}_{3}^{\rm M}\vert},
\end{equation} 
where $\mathbf{R}$ is a 3$\times$3 rotation matrix that rotates a vector by 120$^\circ$ about $\bar{\mathbf{r}}_3^{\rm H}$. $a^*$ and $c^*$ are the effective lattice constants. These effective constants are computed as 
\begin{equation}
a^* = \frac{1}{2}\vert\mathbf{r}_{1}^{\rm M}\vert,\quad
c^* = \frac{3\mathbf{r}_{2}^{\rm H}\cdot(\mathbf{r}_{1}^{\rm M} \times \mathbf{r}_{3}^{\rm M})}{2\vert\mathbf{r}_{1}^{\rm M} \times \mathbf{r}_{3}^{\rm M}\vert}.  
\end{equation}
We find that $a^*=4.609$ and $c^*=11.973$ \AA. 
These effective hexagonal vectors ($\bar{\mathbf{r}}_1^{\rm H}$, $\bar{\mathbf{r}}_2^{\rm H}$, and $\bar{\mathbf{r}}_3^{\rm H}$) are converted to rhombohedral vectors ($\bar{\mathbf{r}}_1^{\rm R}$, $\bar{\mathbf{r}}_2^{\rm R}$, and $\bar{\mathbf{r}}_3^{\rm R}$) \cite{tilley2020} as follows: 
\begin{equation}\label{eqn:rh_rr}
    [\bar{\mathbf{r}}_1^{\rm R},\bar{\mathbf{r}}_2^{\rm R},\bar{\mathbf{r}}_3^{\rm R}] = \frac{1}{3}\begin{bmatrix}
-1.0 & 2.0 & -1.0 \\
1.0 & 1.0 & -2.0 \\
1.0 & 1.0 & 1.0
\end{bmatrix} [\bar{\mathbf{r}}_1^{\rm H},\bar{\mathbf{r}}_2^{\rm H},\bar{\mathbf{r}}_3^{\rm H}],
\end{equation}
These effective rhombohedral vectors ($\bar{\mathbf{r}}_{1}^{\rm R}$, $\bar{\mathbf{r}}_{2}^{\rm R}$, and $\bar{\mathbf{r}}_{3}^{\rm R}$) are used in \sect{sec:result:bisb} to compute elastic stiffness coefficients and lattice constants in the rhombohedral arrangement and compare with those of other Bi-Sb binaries.
In \sect{sec:method:dis_model}, we employ the atomistic structure of monoclinic Bi$_7$Sb$_1$ to model the edge dislocation with $\mathbf{b}=[100]$ because $\mathbf{b}$ lies in the plane spanned by $\mathbf{r}_{2}^{\rm M}$ and $\mathbf{r}_{3}^{\rm M}$ (see right-bottom image in \fig{fig:atom_view}). Therefore, this configuration ensures an infinite dislocation line, as atoms are replicated along $\mathbf{r}_{1}^{\rm M}$ with periodic boundary conditions in the atomistic simulation.
%%%%%%
\subsection{Atomistic model of dislocation in Bi$_7$Sb$_1$}\label{sec:method:dis_model}
\begin{figure}[ht]
    \centering
    \includegraphics[width=0.95\textwidth]{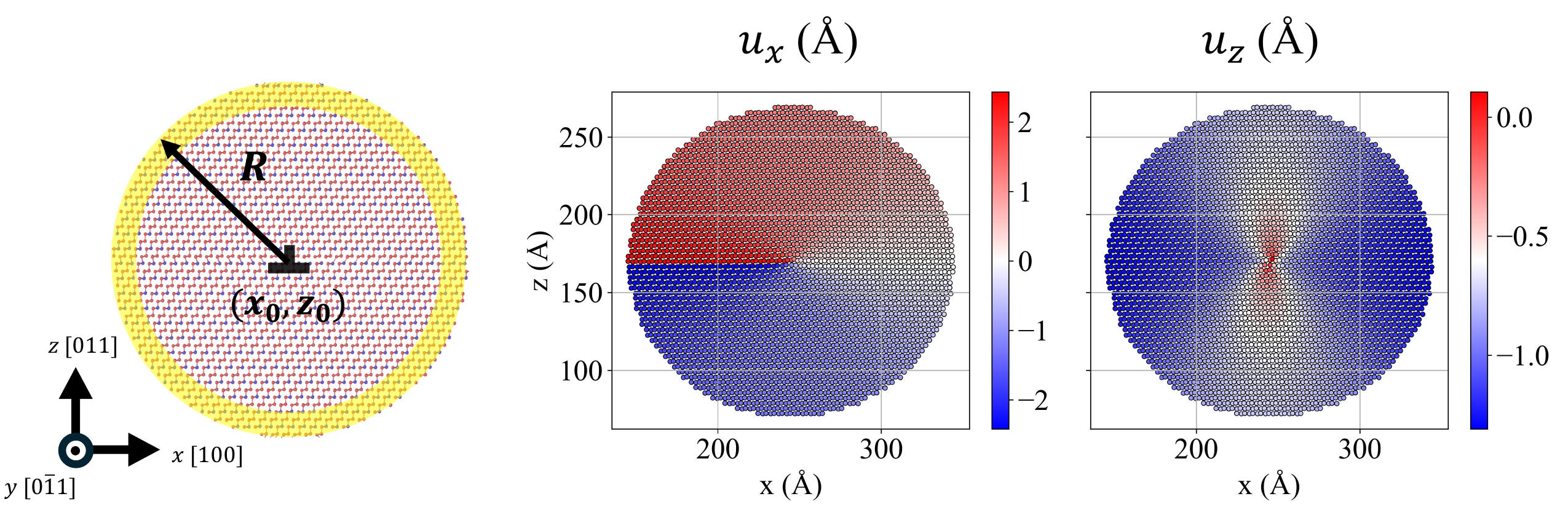}
    \caption{The atomistic structure of a single dislocation ($\mathbf{b}=[100]$) in replicated Bi$_7$Sb$_1$ under fixed boundary conditions. In left figure, atoms in the yellow region are constrained during relaxation. The 2D maps  on the right represent atom displacements along the $x$-axis ($u_x$) and $z$-axis ($u_z$).}
    \label{fig:dis_model}
\end{figure}
\fig{fig:dis_model} shows the atomistic model of a dislocation in Bi$_7$Sb$_1$ and the $u_x$ and $u_z$ displacement fiedls for the initial dislocation generation. The dislocation line lies along the $[100]$ direction ($x$-axis), the $[0\bar{1}1]$ direction aligns with the $y$-axis, and the $[011]$ direction corresponds to the $z$-axis. The simulation box is constructed by replicating the Bi$_7$Sb$_1$ unit cell in a $50\times1\times50$ array. To introduce an edge dislocation, isotropic elastic displacements $u_x$ and $u_z$ are applied to each atom based on its $(x,z)$ position relative to the dislocation core located at $(x_0,z_0)$ (eqns. (3-45) and (3-46) in \cite{lothe2017}).
These displacements are given as: 
\begin{align}
u_x &=\frac{b}{2\pi}\Biggl[\arctan{\Bigl(\frac{\bar{z}}{\bar{x}}\Bigr)}+\frac{\bar{x}\bar{z}}{2(1-\nu)\bar{r}^2}\Biggr], \\
u_z &=-\frac{b}{2\pi}\Biggl[\frac{1-2\nu}{4(1-\nu)}\ln\bigl(\bar{r}^2\bigr)+\frac{\bar{x}^2-\bar{z}^2}{4(1-\nu)\bar{r}^2}\Biggr], 
\end{align}
where $\bar{x}=x-x_0$, $\bar{z}=z-z_0$, $\bar{r}=\sqrt{\bar{x}^2+\bar{z}^2}$, $b = 4.9\;\angstrom$, and Poisson's ratio $\nu$ is $0.3$. After applying $u_x$ and $u_z$, atoms located at a distance $\bar{r}>100\;\angstrom$ from the dislocation core are removed from the simulation. Additionally, in all atomistic simulations, the atoms in the outermost $8\;\angstrom$ of the simulation box (highlighted in yellow in \fig{fig:dis_model}) remain fixed. As noted in \cite{jian2021}, the influence of the boundary on a single dislocation becomes negligible for sufficiently large $R$. We confirm that the dislocation properties do not change once $R$ exceeds $100\;\angstrom$.

\subsection{Active learning algorithm}\label{sec:method:active}
% Algorithm
\begin{figure}[ht]
    \centering
    \includegraphics[width=0.95\textwidth]{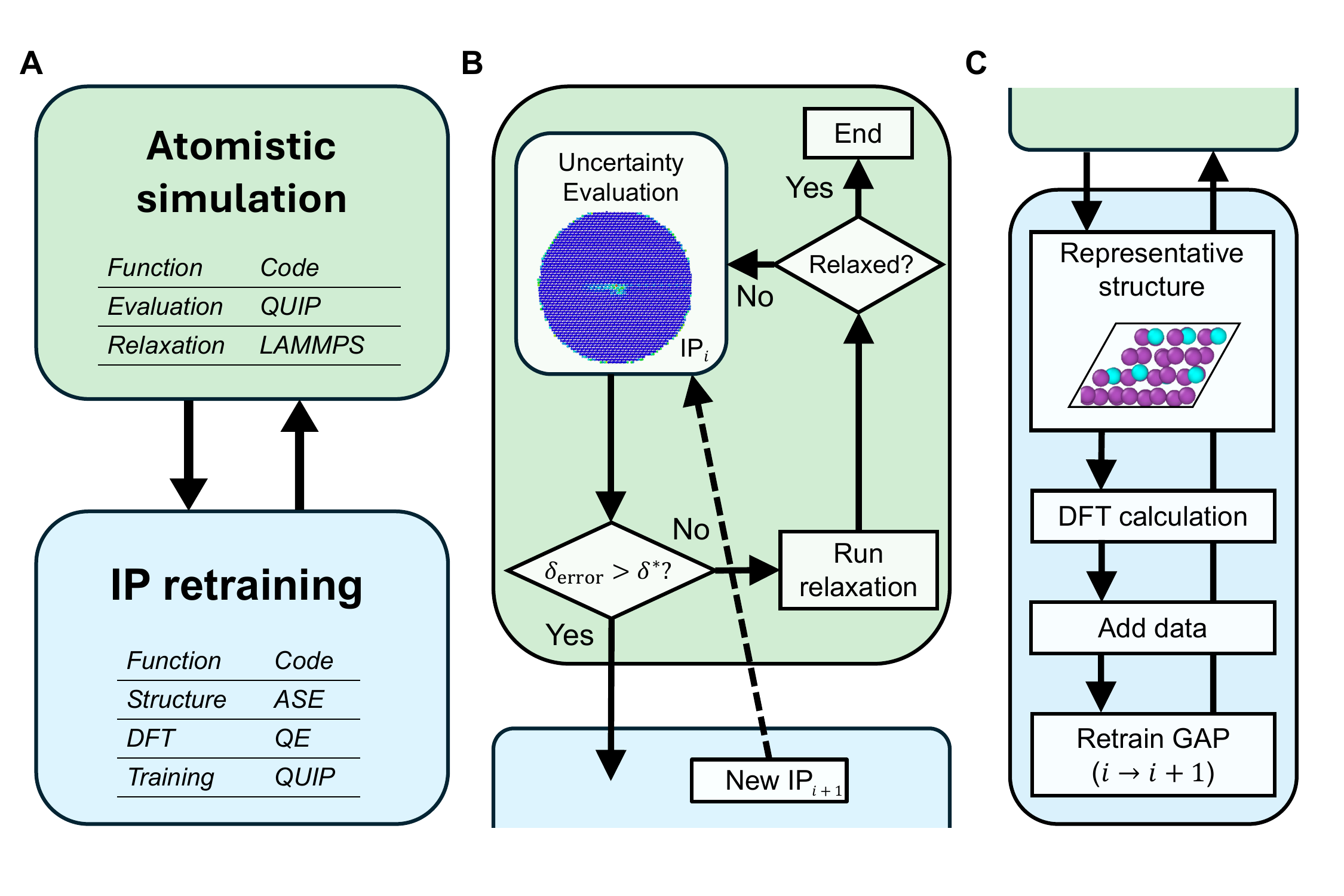}
    \caption{Flowchart of the active learning algorithm. (A) Two main iterative processes, \emph{Atomistic simulation} and \emph{IP retraining}, with the corresponding software packages presented. (B) Subprocess in \emph{Atomistic simulation} for uncertainty evaluation and structural relaxation. (C) Subprocess in \emph{IP retraining}, including the generation of representative structures, DFT calculations, and model retraining.}
    \label{fig:algorithm}
\end{figure}
\fig{fig:algorithm} illustrate the flow chart of our active learning algorithm, which includes two iterative main processes: \emph{Atomistic simulation} and \emph{IP retraining}. In atomistic simulation process, we use a Bi$_7$Sb$_1$ structure with a single dislocation under the fixed boundary condition (see \fig{fig:dis_model}). IP$_0$ is employed as the initial GAP. 
A key advantage of GAP is its capability to predict both target quantities (i.e., forces and energies) and their uncertainties (i.e., variance). Because the force plays a crucial role in determining the relaxation path on a potential energy surface, we focus on the uncertainty of the force components. Specifically, we define the uncertainty of atomic forces ($\delta_{\rm var}$) as
\begin{equation}\label{eqn:force_error}
    \delta_{\rm var} = \sqrt{\text{Var}(f_x)+\text{Var}(f_y)+\text{Var}(f_z)},
\end{equation}
where $\text{Var}(f_x)$, $\text{Var}(f_y)$, and $\text{Var}(f_z)$ are variances of force components along $x$, $y$, and $z$ directions, respectively. 
Whenever $\delta_{\rm var}$ for any atom exceeds a threshold $\delta^*$ of 15 meV/\AA, a representative structure is generated for additional DFT calculation. The threshold of 15 meV/\AA\;is set to ensure that the trained IP reproduces the relaxation path observed during the active learning process (see relaxation path of the active learning and IP$_5$ in \fig{fig:active_min}B). 
The representative structure is constructed as follows. A 2$\times$1$\times$2 supercell is created from the Bi$_7$Sb$_1$ unit cell. The atom with the highest $\delta_{\rm var}$ from the simulation is relocated to the center of this supercell, and its neighboring atoms within the supercell are included. To properly handle periodic boundary conditions, we remove any overlapping atoms: if an atom in the reference cell and its periodic image are within 0.5 \AA\;of each other, one of them is removed. The central atom is maintained at least 9.2 \AA\;from the nearest boundary, ensuring that the interactions with its neighbors in the large simulation are captured effectively in the representative structure. Due to the small size of the 2$\times$1$\times$2 supercell, spurious periodic-boundary interactions may influence the force and energy for the central atom  compared to the full dislocation structure. We checked that force and energy of the center atom vary by less than 2\% when using 3$\times$1$\times$3 supercell. Although a larger cell would improve accuracy, it would also increase computational cost; one representative structure requires 3 hour of wall time with 512 AMD EPYC processors. Once the representative structure is prepared, its energy and force are computed via the DFT calculation. The computed force and energy of the structure are added to the DFT training data set. The GAP potential is then retrained (i.e., $i \rightarrow i+1$) and used to re-evaluate the uncertainty of the structure. If $\delta_{\rm var}$ falls below $\delta^*$, the structure is further relaxed in LAMMPS for $d=1000$ steps using the current IP$_i$. If the norm of force is less than $10^{-6} \rm{eV}/\angstrom$, the simulation terminates. Otherwise, the cycle is repeated until both $\delta_{\rm var}$ and the force norm reach their prescribed criteria. 

%for the atom with highest $\delta_{\rm var}$ with its neighbor atoms is generated for additional DFT calculation (see representative structure in \fig{fig:algorithm}C). This representative structure is created by constructing a supercell (1$\times$2$\times$2 unit cell of Bi$_7$Sb$_1$) and positioning the atom with the highest $\epsilon$ at its center along with neighbor atoms. The central atom lies at least 9.2 $\angstrom$ from the nearest boundary, ensuring that the interactions with its neighbor atoms are captured effectively.\footnote{Due to the small size of the 1$\times$2$\times$2 supercell, spurious periodic-boundary interactions may influence the force and energy for the central atom  compared to the full dislocation structure. We checked that force and energy of the center atom vary by less than 2\% when using 1$\times$3$\times$3. Although a larger cell would improve accuracy, it would also increase computational cost; one representative structure  requires 3 hour of wall time with 512 AMD EPYC processors} 
\section{Result}
\subsection{Structural and elastic Properties of Bi$_{1-x}$Sb$_{x}$}\label{sec:result:bisb}
\begin{figure}[ht]
    \centering
    \includegraphics[width=0.7\textwidth]{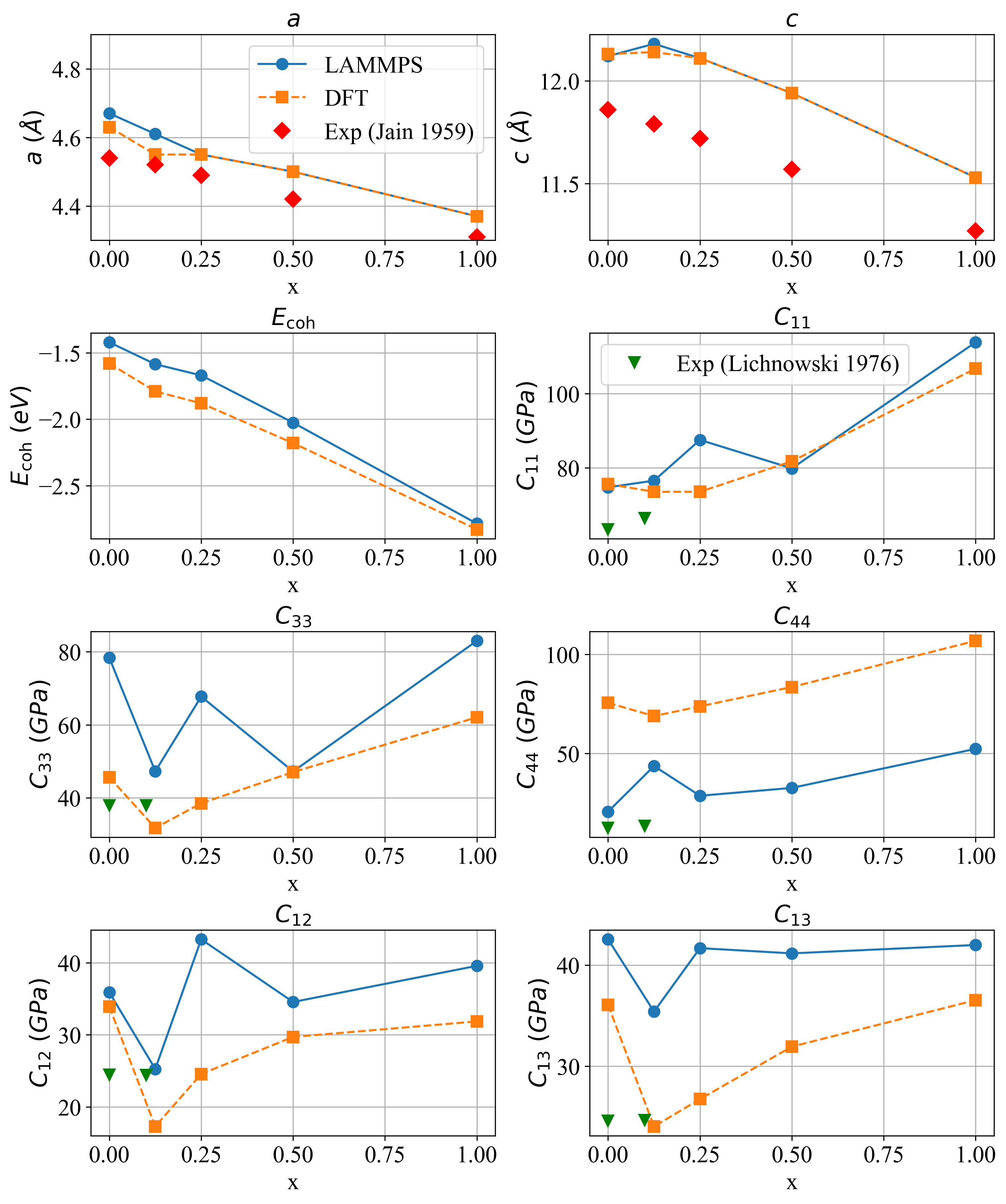}
    \caption{Structural and elastic properties change of Bi$_{1-x}$Sb$_x$ as a function of Sb concentration. Both DFT and GAP results are shown, with experimental data (\cite{jain1959} and \cite{lichnowski1976}) provided where available.}
    \label{fig:prop_sb}
\end{figure}
\fig{fig:prop_sb} displays the computed structural and elastic properties of Bi$_{1-x}$Sb$_x$ across a range of Sb composition ($x$), obtained from both our GAP (IP$_5$) and DFT calculations. For comparison, experimental data from \cite{jain1959} and \cite{lichnowski1976} are also presented. The GAP potential successfully reproduces the key trends observed in the DFT results, particularly for the structural parameters. As the Sb composition ($x$) increases from 0 (pure Bi) to 1 (pure Sb), the in-plane lattice parameter ($a$) decreases from 4.63 Å (GAP) and 4.67 Å (DFT) to 4.367 Å (GAP) and 4.37 Å (DFT), respectively. Similarly, the out-of-plane lattice parameter ($c$) decreases from 12.13 Å (GAP) and 12.23 Å (DFT) to 11.53 Å (GAP) and 11.54 Å (DFT). These linear and monotonic trends align well with experimental observations, although the absolute values of $a$ and $c$ exhibit offsets compared to the experimental data.
Similar to the behavior observed in the lattice parameters ($a$ and $c$), the cohesive energy (E$_{\rm coh}$) decreases monotonically with increasing Sb concentration ($x$). Specifically, E$_{\rm coh}$ shifts from $-1.42$ eV (GAP) and $-1.58$ eV (DFT) at $x=0$ (pure Bi) to $-2.78$ eV (GAP) and $-2.83$ eV (DFT) at $x=1$ (pure Sb). This continuous decrease indicates that the bonding in Bi$_{1-x}$Sb$_{x}$ becomes progressively stronger as the Sb concentration increases.

The mechanical stiffness of Bi$_{1-x}$Sb$_x$ (\fig{fig:prop_sb}) strongly depends on Sb concentration ($x$). Both DFT and GAP calculations indicates that the in-plane stiffness constant (C$_{11}$) increases with Sb concentration while showing a minimum at $x=0.125$ (Bi$_7$Sb$_1$) before rising. Although GAP reproduces C$_{11}$ well, discrepancies emerge in the other elastic constants. These differences highlight the challenge of training a single potential to capture all properties across the BiSb binary system (Bi, Bi$_7$Sb$_1$, Bi$_3$Sb$_1$, BiSb, and Sb). While expanding the training dataset and adding descriptors could reduce these discrepancies, such modifications would significantly increase computational costs for large atomic systems. Given that our primary focus is on the dislocation behavior in Bi$_7$Sb$_1$, where dislocation-induced topological states emerge, we have optimized our GAP specifically to reproduce the DFT results for Bi$_7$Sb$_1$.
%%%%%%%%%%%%%%%%
\subsection{Relaxation of dislocation using active learning}\label{sec:result:relaxation}
\begin{figure}[ht]
    \centering
    \includegraphics[width=\textwidth]{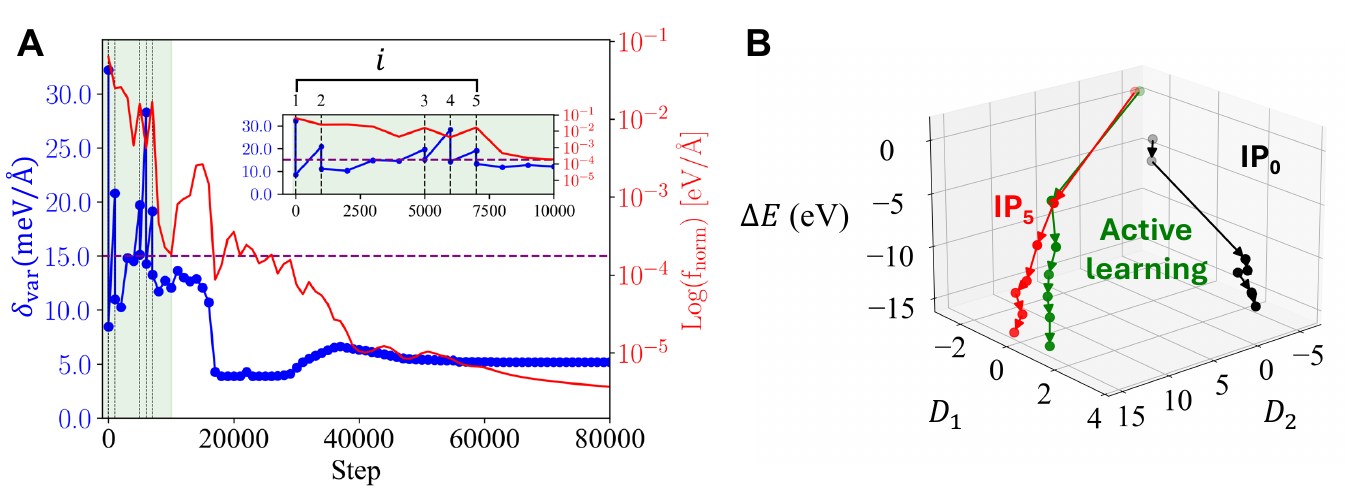}
    \caption{Dislocation relaxation in Bi$_7$Sb$_1$ using the active learning. (A) Evolution of force error and force norm ($f_{\rm norm}$) during the relaxation via the active learning. The black dotted line indicates the drop in force error following the retraining of the interatomic potential (IP$_i\rightarrow$\,IP$_{i+1}$), while the green-shaded region highlights the retraining phase (a zoomed-in view is provided in the inset). (B) Two-dimensional UMAP projection of the local atomic environments near the dislocation core. Relaxation trajectories for different IPs are shown: IP$_5$ (red), active learning relaxation (green), and IP$_0$ (black), depicted with arrows and circles. $\Delta E$ denotes the energy difference of each configuration relative to the initial structure as predicted by IP$_5$.}
    \label{fig:active_min}
\end{figure}
\fig{fig:active_min} shows the relaxation process, highlighting both the evolution of the $\delta_{\rm var}$ (\fig{fig:active_min}A) and structure with different IPs (\fig{fig:active_min}B). In \fig{fig:active_min}A, the evolution the uncertainty of the atomic force ($\delta_{\rm var}$ defined in \eqn{eqn:force_error}) is presented as the system relaxes under the active learning (see \fig{fig:algorithm} in \sect{sec:method:active} for details). $\delta_{\rm var}$ is evaluated every 1000 simulation steps, and whenever it exceeds the threshold of 15 meV/\AA, a representative structure is generated. The force and energy of this representative structure are computed using DFT, and GAP is then retrained with an updated training set (IP$_{i+1}$). In \fig{fig:active_min}A, this retraining is shown as a sudden drop in the $\delta_{\rm var}$ at the same step (marked as vertical black dotted line). Consequently, the active learning process effectively maintains the $\delta_{\rm var}$ (uncertainty of the atomic force) below 15 meV/\AA\;throughout the entire relaxation process. 
In the early stage of relaxation, where the force norm is high, the structure experiences significant changes, leading to the elevated uncertainty. During this stage (highlighted in green region in \fig{fig:active_min}A) five representative structures are generated and incorporated into the training set. With each retraining step, the overall error is maintained below 15 meV/\AA, eventually stabilizing around 5 meV/\AA\;after 20,000 steps. Beyond this point, the structure exhibits minor changes, and no further retraining is needed due to $\delta_{\rm var} < \delta^*$.  

\fig{fig:active_min}B further illustrates the evolution of the atomic configuration near the dislocation core (within 20 \AA) under different interatomic potentials (IPs) and during the active learning process. To evaluate structural differences, we applied the Uniform Manifold Approximation and Projection (UMAP) method to the local atomic environments at the dislocation core using two distinct projections: UMAP $D_1$ and UMAP $D_2$ (see Supporting Information for details) \cite{McInnes2018}. In \fig{fig:active_min}B, each point connected by consecutive arrow represents $D_1$ and $D_2$ projections of the dislocation core structure obtained during relaxation of 3000 steps performed by IP$_5$ (red) and IP$_0$ (black). The green points are obtained from structural data collected every 3000 steps during the active learning process. Here, $\Delta E$ denotes the energy change of each configuration relative to the energy of initial structure predicted by IP$_5$. During the first 3000 relaxation steps, the structures relaxed by IP$_5$ closely match with those obtained from the active learning process. However, beyond the first point (3000 steps), the two trajectories begin to diverge slightly. In contrast, the relaxed structures using IP$_0$ lose accuracy compared to those from IP$_5$ and active learning process at the initial point; this is because IP$_0$ lacks of prior information associated with the dislocation core, leading to the significant divergence as the relaxation proceeds. 
Overall, these divergences in structural evolution indicate two key points: (1) different interatomic potentials can follow distinct relaxation pathways, and (2) IP$_5$, when actively retrained on representative dislocation core configurations, reproduces a relaxation pathway similar to that observed with the active learning approach.
The divergence between structures produced by IP$_5$ and those obtained through the active learning process decreases when $\delta^*$ is set to a low value (currently 15 meV/\AA), though this reduction comes at the cost of increased computational time. In the following sections, we employ IP$_5$ to investigate dislocation mechanics in Bi$_7$Sb$_1$, specifically in the Peierls stress calculations (\sect{sec:result:peierls}) and the compression simulations (\sect{sec:result:compression}).
%%%%
\subsection{Peierls Stress Calculation}\label{sec:result:peierls}
\begin{figure}[ht]
    \centering
        \includegraphics[width=\textwidth]{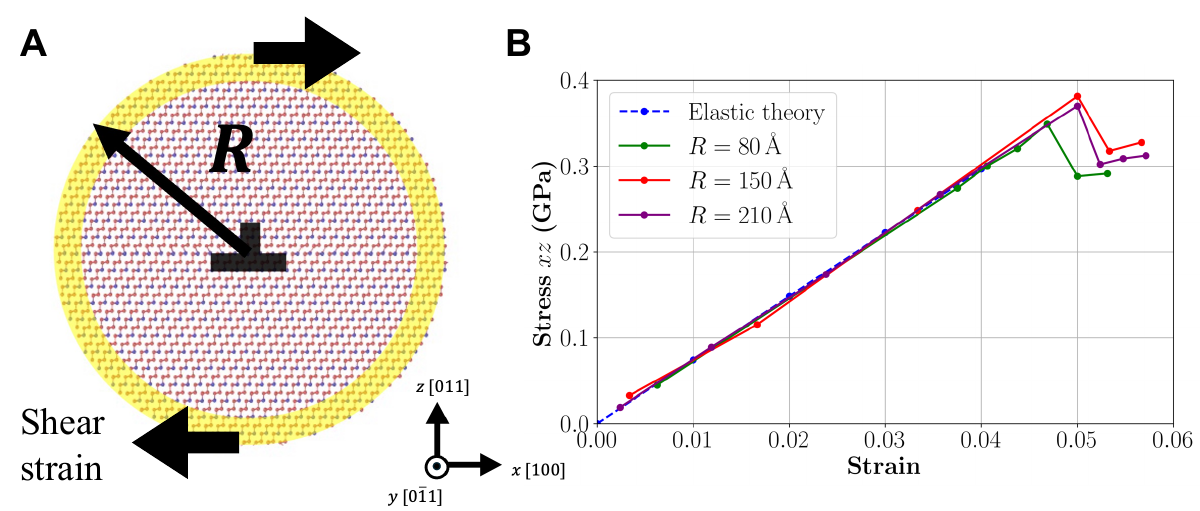}
        \caption{(A) Bi$_7$Sb$_1$ structure with a dislocation at the center. 
        Fixed boundary conditions are applied. (B) Shear stress as a function of strain. The blue dotted line indicates the linear elastic response with $E = 45.88$ GPa.}
    \label{fig:peirels_stress}
\end{figure}
\fig{fig:peirels_stress} illustrates the calculation of the Peierls stress (the minimum shear stress required to displace a dislocation). In \fig{fig:peirels_stress}A, the dislocation with $\mathbf{b}=[100]$ in the Bi$_7$Sb$_1$ structure (the same model described in \sect{sec:method:dis_model}) is shown. Two black arrows represent the direction of the applied $xz$ shear strain to induce the motion of the dislocation. As the structure is gradually strained, the applied atomic stress is computed at each increment in strain after the system is relaxed. The stress initially increases linearly, as expected for elastic deformation (see blue dotted line in \fig{fig:peirels_stress}B); however, once the dislocation overcomes the Peierls barrier and start to move, a sudden drop in stress is observed. The peak stress just before this drop is recorded as the Peierls stress.

In \fig{fig:peirels_stress}B, we explore the influence of system size ($R$) on the Peierls stress. Simulations are performed for increasing values of $R$ (80, 150, and 210 \AA).
Although other boundary conditions, such as fixed boundary square, dislocation dipole and quadrupole are available, \cite{jian2021} demonstrates that for sufficiently large systems the Peierls stress remains essentially unchanged for different boundary conditions.
In \fig{fig:peirels_stress}B, the $xz$ shear stress is shown as a function of strain, with the blue dotted line representing the prediction from linear elastic theory ($E = 45.88$ GPa; see Supplementary Materials for the computation of $E$). For all system sizes, the stress increases linearly with strain during the elastic regime. In the smallest system ($R = 80\;\angstrom$), the stress drops at a strain of 0.046, yielding a Peierls stress of 0.35 GPa. In contrast, for larger systems ($R = 150$ and $210\;\angstrom$), the stress drop occurs at a strain of 0.05, corresponding to a Peierls stress of 0.39 GPa, showing convergence of the Peierls stress for $R \geq 150\;\angstrom$.
Throughout the straining process, the force error remains below 15 meV/\AA. Although the force error exceeds this threshold value when the dislocation moves, our primary objective is to determine the maximum stress just before the dislocation moves. Therefore, we do not retrain the potential to capture forces and energies after dislocation moves. The fact that the maximum uncertainty remains within our set threshold confirms that the strain simulation stays close to the configurations represented in the training data.
%%%%
\subsection{Dislocation formation in Bi$_7$Sb$_1$ under $[\bar{2}11]$ compression}\label{sec:result:compression}
\begin{figure}[ht]
    \centering
    \includegraphics[width=\textwidth]{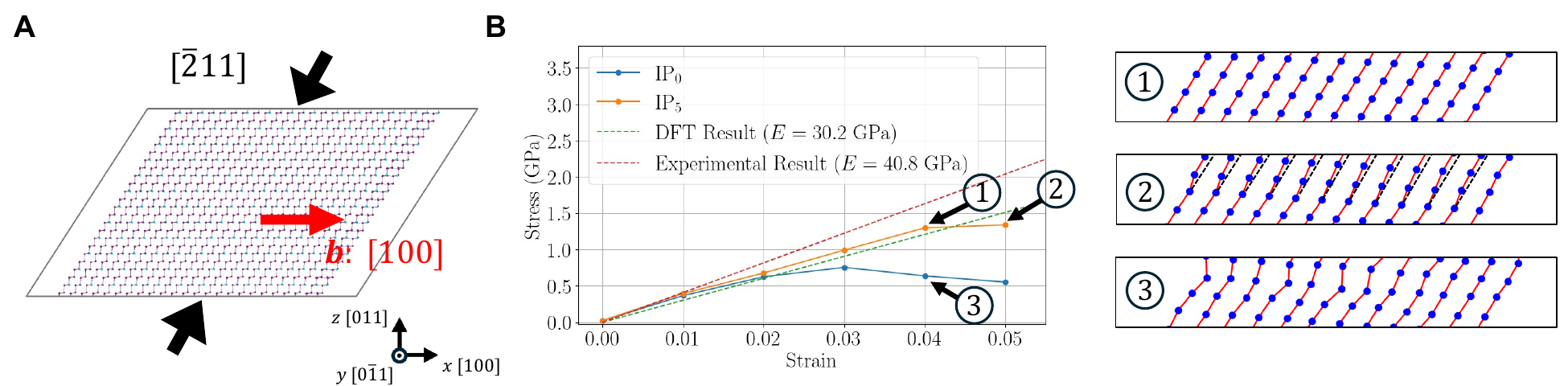}
    \caption{Dislocation formation under $[\bar{2}11]$ compression. (A) Schematic representation of $[\bar{2}11]$ with directional information. Red arrow represent Burgers vector of $\mathbf{b}=[100]$. (B) Stress-strain curve for compression with IP$_0$ and IP$_5$ configurations compared to the linear elastic response derived from DFT calculations and experimental results \cite{lichnowski1976}. Atomic configurations of Sb atoms at states $\textcircled{1}$, $\textcircled{2}$, and $\textcircled{3}$ are shown on the right panel. The black dotted line in configuration $\textcircled{2}$ represents a reference line from the linear configuration $\textcircled{1}$, illustrating the shearing displacement of atoms during dislocation formation.}
    \label{fig:compression}
\end{figure}
To investigate dislocation formation under compressive loading, we impose uniaxial strain along the $[\bar{2}11]$ direction on a Bi$_7$Sb$_1$ structure. \fig{fig:compression}A illustrates the simulation setup prepared for the $[\bar{2}11]$ compression. The Bi$_7$Sb$_1$ unit cell is replicated into a $16\times1\times16$ supercell, and a free surface along $x$-direction ($[100]$) is created by enlarging the simulation box. This configuration facilitates slip between atomic layers, ultimately leading to dislocation formation.
The strain along $[\bar{2}11]$ is applied via a deformation gradient $\mathbf{F}$ on lattice vectors, assuming a small strain regime:
\begin{equation}
    \mathbf{F} = \mathbf{I}+\epsilon \mathbf{n} \otimes \mathbf{n},
\end{equation}
where $\mathbf{I}$ is the identity tensor, $\epsilon$ is the applied strain, and 
\begin{equation}
\mathbf{n}=-2\bar{\mathbf{r}}_1^{\mathbf R}+\bar{\mathbf{r}}_2^{\mathbf R}+\bar{\mathbf{r}}_3^{\mathbf R} 
\end{equation}
defines the [$\bar{2}11$] direction in Cartesian coordinate. The strain is incrementally increased from 0.0 up to 0.05, and at each step the structure is relaxed using the FIRE minimization algorithm.
The corresponding normal stress along $[\bar{2}11]$, denoted as $\sigma_{[\bar{2}11]}$, is computed via ($\sigma_{[\bar{2}11]}$), using
\begin{equation}
    \sigma_{[\bar{2}11]} = \boldsymbol{\sigma}\mathbf{n}
\end{equation}
where $\boldsymbol{\sigma}$ is a stress tensor of the entire system obtained by LAMMPS. 
\fig{fig:compression}B presents the measured stress for both IP$_0$ and IP$_5$, along with linear fits obtained from linear elasticity using elastic stiffness from DFT and experimental data \cite{lichnowski1976}. 
The effecitve young's modulus for $[\bar{2}11]$ direction for DFT and experimental data is computed following the approach described in \cite{nye1985} (the relevant equations are provided in the Supplementary Materials). 
\fig{fig:compression}B presents that the stress–strain curves for both IP$_0$ and IP$_5$ display response, which agrees closely with the DFT (red dotted line) and experimental (green dotted line) predictions. In the case of IP$_5$, the linear elastic response persists up to a strain of 0.04, with structure  $\textcircled{1}$ showing uniform deformation of Sb atoms along $[\bar{2}11]$. At a strain of 0.05, this linear response breaks down, and a pronounced slip is observed (structure $\textcircled{2}$), indicating the slip and the formation of a dislocation, as highlighted by the dotted black line.
In contrast, IP$_0$ shows that the stress-strain relationship deviates from linearity at a strain of 0.03, indicating early plastic deformation compared to that of IP$_5$. The arrangement of Sb atoms at this strain (structure $\textcircled{3}$) reveals abnormal deformation, due to the absence of prior information about slip or dislocation in IP$_0$. These results demonstrate that IP$_5$ not only captures the linear elastic behavior but also successfully reproduces dislocation core generation via slip, as facilitated by the active learning process.\footnote{During the relaxation of strained structure with IP$_5$, we also confirm that the maximum force uncertainty does not exceed threshold criteria $\delta^*$ (15 meV\AA).}
Furthermore, variations in the lateral width of the simulation cell (along the $x$ direction) indicate that increasing the width beyond the current dimensions does not affect the stress measurements or the overall stress–strain response.

\section{Conclusion}
We develop an IP for Bi-Sb binaries, with a particular focus on the topological insulator Bi$_7$Sb$_1$ and its dislocation. Our active learning framework combines DFT with classical atomistic simulations. In this approach, the Bi$_7$Sb$_1$ structure containing a dislocation is relaxed using GAP and re-trained by DFT calculation of a representative structure of dislocation core whenever the force uncertainty exceeds threshold. Five representative structures were generated to effectively capture the force and energy variations near the dislocation core while maintaining a low force error during relaxation. UMAP analysis shows that the trained IP reliably reproduces the structural evolution observed during active learning in contrast to the initial IP which deviated significantly. The final IP was then applied to calculate dislocation properties, including the Peierls stress and dislocation generation under compression. We observed convergence of the Peierls stress with increasing system size and note that dislocation generation by slip during compression was successfully captured.

Despite incorporating active learning to capture dislocation mechanics, our approach currently relies on a single relaxation pathway on one energy surface, which limits its broader applicability. This constraint diminishes the robustness of IP when predicting structures outside the training dataset, such as those found in finite-temperature simulations with stochastic vibrations. A challenge in interatomic potential development is assembling a dataset that is both representative and geometrically diverse that can accurately capture the mechanics of material. Future research may address these limitations by employing multiple existing interatomic potentials to perform parallel relaxations on the same material. Such a strategy would enable the collection of a wider variety of structural data, facilitating the construction of a more diverse initial dataset and ultimately enhancing the reliability of trained IP.

Furthermore, while equivariant neural network models, such as graph neural networks, enable efficient simulations with relatively small datasets and high computational speed, their current inability to efficiently provide uncertainty estimates limits their practical application in active learning workflows. Developing robust uncertainty quantification methods for these models could significantly accelerate the development of interatomic potentials.

\bibliography{main}
\end{document}